\documentclass[pra, aps, twocolumn, groupedaddress, showpacs, superscriptaddress]{revtex4-2}
\usepackage{amssymb,amsmath,amsthm,color,graphicx,times,graphicx}
\usepackage[caption=false]{subfig}
\usepackage{hyperref}
\usepackage{enumerate}
\usepackage{graphicx}
\usepackage{dsfont}
\usepackage{times}
\usepackage{bbold}
\usepackage{color}
\usepackage{orcidlink}
\usepackage{hyperref}

\hypersetup{
colorlinks=true,
linkcolor=blue,
filecolor=blue,
urlcolor=blue,
citecolor=violet,
pdfauthor={ F. Aljuaydi, S. N. almutairi, and   A.-B. A. Mohamed},
pdftitle={},
pdfcreator={F. Aljuaydi},
}
\providecommand{\openone}{\leavevmode\hbox{\small1\kern-4.3pt\normalsize1}}

\theoremstyle{plain}

\theoremstyle{definition}

\input{tcilatex}
\begin{document}

\title{Generation and robustness of non-local correlations induced by Heisenberg XYZ and   intrinsic decoherence models:   $(x,y)$-spin-orbit interactions and $x$-    magnetic field     }

\author{F. Aljuaydi}\email{Corresponding author: f.aljuaydi@psau.edu.sa}

\author{S. N. Almutairi}
\address{Department of Mathematics, College of Science and Humanities, Prince Sattam bin Abdulaziz University, Al Kharj 11942, Saudi Arabia}

\author{A.-B. A. Mohamed} 
\address{Department of Mathematics, College of Science and Humanities, Prince Sattam bin Abdulaziz University, Al Kharj 11942, Saudi Arabia}
\address{ Department of Mathematics,  Faculty of Science, Assiut University, Assiut, Egypt}

\begin{abstract}
In this work, the Milburn intrinsic decoherence model is used to investigate the role of spin-spin Heisenberg-XYZ interaction supported by spin-orbit Dzyaloshinsky–Moriya (DM) interactions of $x$ and $y$-directions together in the non-local correlation (NLC) dynamics of  Local quantum Fisher information (LQFI), local quantum uncertainty (LQU), and Log-negativity's entanglement.  The two-qubit-Heisenberg-XYZ (non-X)-states' non-local correlation generations are explored under the effects of the uniformity and the inhomogeneity of an applied   $x$-direction external inhomogeneous magnetic field (EIMF). 
 Our meticulous exploration of the obtained results shows that the spin-spin  Heisenberg XYZ and  $x,y$-spin-orbit interactions have a high capability to raise non-local correlations in the presence of a weak external magnetic field.
 The raised non-local correlation can be improved by strengthening the spin-spin and $x,y$-spin-orbit  interactions and increasing the  EIMF's inhomogeneity and uniformity.    Non-local correlation oscillations' amplitudes and fluctuations are increased.
The degradations of the NLCs' generations in the presence of intrinsic decoherence (NLCs' robustness against intrinsic decoherence)  can be decreased by strengthening the spin-spin interactions. They can be increased by increasing the intensities of     $x,y$-spin-orbit interactions as well as increasing the  EIMF's inhomogeneity and uniformity.
\end{abstract}

\maketitle

\section{Introduction}
Among the numerous quantum systems proposed to implement
 quantum  information and computation \cite{Horod-1,Nielsen2000},   superconducting circuits, trapped ions, and semiconductor quantum dots are essential techniques for realizing quantum bits (qubits).
Based on electron spins trapped in quantum dots,   a quantum
computer protocol has been initially proposed \cite{IntrodC-1,IntrodC-2,IntrodC-2b}, the electron having a spin of 1/2 is the simplest natural qubit. 
 Recently,  quantum computation (as a single-spin-qubit geometric gate) with electron spins (single-spin-qubit geometric gate) has been  realized  in   quantum dots \cite{IntrodC-3,IntrodC-4}. 
Due to tunneling the electrons from one to the other, the spin-spin coupling and spin-orbit coupling of the interaction  between two qubits can be realized by considering a two-qubit system represented by two coupled quantum dots' two electrons. 
Therefore, Heisenberg XYZ models describing spin-spin interactions are among of the important proposed qubit systems.  
Two qubit Heisenberg XYZ models have been realized in different systems, including bosonic atoms inside an optical lattice \cite{Heisenberg-2013}, trapped ions \cite{Heisenberg-2017} and superconductor systems \cite{Heisenberg-2007}, and linear molecules   \cite{Heisenberg-2022}. Heisenberg XYZ models have been updated to include
spin-orbit interactions  \cite{Hei-6, Hei-7}  with the first order of  SO coupling ( that is known by  Dzyaloshinsky–Moriya interactions \cite{HAM-9} (realizing  by   an antisymmetric superexchange La$_2$CuO$_4$ interaction   \cite{Ksa-2}), the second order of  SO coupling (Kaplan-Shekhtman-Entin-Wohlman-Aharony  interaction \cite{HAM-10}). Spin-1/2 Heisenberg XYZ models also have been updated to include dipole-dipole interaction \cite{dipole-10}, and inhomogeneous external magnetic fields (IEMFs) \cite{Mag-10a,Mag-10}.
\\ \indent
Exploring two-qubit information dynamics in different proposed qubit systems to two-qubit resources, relating to different types of nonlocal correlations  (as entanglement, quantum discored, ...), is one of the most required research fields in implementing quantum information and computation \cite{Bellac2006}. 
 Quantum entanglement (QE)  (realizing by quantifiers' entropy \cite{Entropy-1}, concurrence \cite{Conc}, negativity, and log-negativity \cite{Mn17}, ...) is an important type of qubits' nonlocal correlations (NLCs) \cite{Concy-1,Concy-2} and applications have an important role in quantum information fields. Where  QE has a wide range of applications in implementing quantum computation, teleportation \cite{ENT-1,ENT-2}, quantum optical memory \cite{ENT-3}, and quantum key distribution \cite{ENT-4}. 
 After implementing quantum discord as   another type of qubits' NLCs beyond entanglement 
\cite{QDiscord-1},  several NLCs' quantifiers have been introduced to address other  NLCs \cite{GM-1,GM-2} by using     Wigner–Yanase (WY) skew information \cite{U0} and quantum Fisher information (QFI) \cite{QFI6}. Where, WY-skew-information minimization (local quantum uncertainty \cite{U1} LQU) and the WY-skew-information maximization (uncertainty-induced nonlocality \cite{U2}) have been introduced to quantify other  NLCs beyond entanglement.  
Also, the minimization of QFI (local quantum Fisher information, LQFI) was used to  implementing other qubits' NLCs \cite{QFI7,Kim2018}.  LQU has a direct connection to LQFI \cite{Helstrom1979,Paris2009}, establishing more two-qubit NLCs in several proposed qubit systems \cite{Fishp2021,Fishp2021b}: as hybrid-spin systems (under random noise \cite{Ben2022C} and intrinsic decoherence \cite{Ben2023}), two-coupled double quantum dots \cite{LQFU-12022},     the mixed-spin Heisenberg   \cite{LQFU-12024},     Heisenberg XXX system \cite{LQFU-12022v}.
 \\ \indent
The information dynamics of the two-spin Heisenberg XYZ states
have been investigated,   by using   Milburn intrinsic decoherence model  \cite{Ms1}, of entanglement teleportation   based on the Heisenberg XYZ chain \cite{Ms-Us-1,Ms-Us-2}, Fisher of  Heisenberg XXX states'LQFI beyond IEMF effects \cite{LQFU-12023}, quantum correlations of concurrence and LUQ \cite{Ms-Us-3}. 
The previous works have focused on exploring the time evolution of the two-spin Heisenberg-XYZ states' NLCs with limited conditions on the spin-spin and spin-orbit interactions, and the applied magnetic fields, to ensure residing two-qubits    X-states \cite{Addq1,Addq2,Addq3,Addq4,Addq5,Addq6,QIfom-4,QIfom-3a,UA5}.
Therefore, by using the  Milburn intrinsic decoherence and Heisenberg XYZ models are used to investigate the non-local correlation dynamics of  LQFI,  LQU, and log-negativity (LN) for general two-qubit-Heisenberg-XYZ (non-X)-states, inducing by other specific conditions on the spin-spin and spin-orbit interactions, as well as the applied magnetic fields.
\\ \indent
The manuscript structure is prepared to include the Milburn intrinsic decoherence equation including the Heisenberg XYZ model and its solution in   Sec. (\ref{SEC-2}). But in Sec. (\ref{SEC-3}), we introduce the definition of the NLCs' quantifiers of LQFI,  LQU, and  LN.  Sec. (\ref {SEC-4}) presents the outcomes of the dependence of the NLCs' quantifiers on the physical parameters.  
  Our conclusions are provided in Sec. (\ref{SEC-5}).
 
 \section{The    Heisenberg spin  model}\label{SEC-2}
Here, Milburn intrinsic decoherence model and Heisenberg XYZ model are used to examine the embedded capabilities in spin-spin interaction and spin-orbit interaction (that describes   Dzyaloshinsky–Moriya (DM) $x,y$-interactions  with the first order of  SO couplings   $D_{x}$ and $D_{x}$) to generate essential two SO-qubits' nonlocal correlations (NCs)  under the effects of the uniformity  $B_{m}$ and the inhomogeneity   $b_{m}$  of applied external inhomogeneous magnetic field (EIMF).
For two spin-qubits (each $k$-qubit  ($k=A,B$) is described by upper $|1_{k}\rangle$ and  lower $|1_{k}\rangle$ states), the Hamiltonian of the system is written as
 \begin{eqnarray} \label{Q1}
\!\!\!\!\!\hat{H}=\!\!\sum_{\alpha=x,y,z}  J_{\alpha}\hat{\sigma}^{\alpha}_{A}\hat{\sigma}^{\alpha}_{B}+\sum_{k=A,B} \vec{B}_{k}.\vec{\sigma}_{k}
+\vec{D}.(\vec{\sigma}_{A}\times\vec{\sigma}_{B}).
\end{eqnarray}%
$\vec{\sigma}_{k}=(\hat{\sigma}_{k}^{x}, \hat{\sigma}_{k}^{y},\hat{\sigma}_{k}^{z})$ with $\hat{\sigma}_{k}^{x,y,z}$ represent the $k$-qubit Pauli matrices. $\vec{B}_{k}=(B^{x}_{k},B^{y}_{k},B^{z}_{k})$ is the vector of the  external magnetic field applying on $k$-spin,   
$\vec{B}_{k}.\vec{\sigma}_{k}=B^{x}_{k}\hat{\sigma}_{k}^{x}+B^{y}_{k}\hat{\sigma}_{k}^{y}
+B^{z}_{k}\hat{\sigma}_{k}^{z}$. 
In our work, we consider that the  EIMF is applied only in the  $x$-direction: $\vec{B}_{k}=(B^{x}_{k},0,0)$,  $B^{x}_{A}=B_{m}+b_{m}$, and  $B^{x}_{B}=B_{m}-b_{m}$. 
$\vec{D}=(D_{x},D_{y},D_{z})$   is the spin-orbit/DM  interaction vector. Therefore, we have
$\vec{D}.(\vec{\sigma}_{A}\times\vec{\sigma}_{B})=D_{x}\hat{C}_{x}+D_{y}\hat{C}_{y}
+D_{z}\hat{C}_{z}$ 
with $\hat{C}_{\alpha}=\hat{\sigma}_{A}^{\alpha+1}\hat{\sigma}_{B}^{\alpha+2}
-\hat{\sigma}_{A}^{\alpha+2}\hat{\sigma}_{B}^{\alpha+1} (\alpha=x,y,z)$. 
\\ \indent
Here, we take only the $x,y$-spin-orbit interactions: $\vec{D}=(D_{x},D_{y},0)$.  
The capacitive spin-spin and $x,y$-spin-orbit interactions, under the effects of the EIMF characteristics, can be used to build two-spin-qubit correlations. The considered Hamiltonian is written as
 \begin{eqnarray} \label{Q1}
 \hat{H}&=&\sum_{i=x,y,z}  J_{i}\hat{\sigma}^{i}_{A}\hat{\sigma}^{i}_{B}+\sum_{i=x,y}D_{i}\hat{C}_{i}\nonumber\\&&\qquad\quad+ (B_{m}+b_{m})\hat{\sigma}^{x}_{A}+ (B_{m}-b_{m})\hat{\sigma}^{x}_{B}.
\end{eqnarray}%
\\ \indent
The  time evilution  of the two spin-qubits' nonlocal correlations (NCs)  will be explored by using   Milburn intrinsic decoherence model   \cite{Ms1}, which is given by
\begin{equation}\label{E9}
	\frac{d}{d t}  \hat{M}(t)=-i[\hat{H},\hat{M}]-\frac{\gamma}{2}[\hat{H},[\hat{H},\hat{M}]],
\end{equation}
$\hat{M}(t)$  is the density matrix of the generated two-spin-qubits  state. $\gamma$ is the intrinsic spin-spin  decoherence  (ISSD) coupling.
\\ \indent
Here, the  two-spin-qubits eigenvalues $V_{k}$ ($k=1, 2, 3, 4$) and the   eigenstates $|V_{k}\rangle$ of the  Hamiltonian of Eq. (\ref{Q1})  will be  calculated, numerically. And hence, in the two-spin-qubits  basis: $\{|1_{A}1_{B}\rangle, |1_{A}0_{B}\rangle,  |0_{A}1_{B}\rangle, |0_{A}0_{B}\rangle\}$, the   two-spin-qubits  state dynamics  can obtained numerically by using   the  solution of Eq. (\ref{E9})  giving by 
\begin{equation}\label{E4}
	\hat{M}(t)= \!\! \sum^{4}_{m, n=1} \!\! U_{mn}(t) \,S_{mn}(t)
	\,\langle V_{m}|\hat{M}(0)|V_{n}\rangle\,
	|V_{m}\rangle\langle V_{n}|.
\end{equation}
The   unitary interaction  $U_{mn}(t)$ and  the  ISSD coupling $S_{mn}(t)$ effects are controlled by the following terms:
\begin{eqnarray}\label{E4XX}
	U_{mn}(t)&=&
	e^{-i(V_{m}-V_{n})t}, \nonumber\\
 S_{mn}(t)&=&e^{ -\frac{\gamma}{2}
		(V_{m}-V_{n})^{2}t}.
\end{eqnarray}
The Eq.(\ref{E4}) is used to calculate and explore, numerically,  the dynamics of the nonlocal correlations residing within the two-spin-qubits states' Heisenberg XYZ model under the effects of the $x,y$-spin-orbit interactions (spin-orbit interactions acting along the $x-$ and $y-$ directions)  and an applied external magnetic field applying along $x$-direction.   
\section{Non-local correlation  (NLC) quantifiers}\label{SEC-3}
Here, the two-spin-qubits' NLCs  will be measured by the following  LQFI, LQU, and     logarithmic negativity (LN):
\begin{itemize}
\item \textbf{ LQFI}\newline
LQFI can be used   as a two-spin-Heisenberg-XYZ correlation quantifier beyond entanglement, which is recently introduced  as another correlation type. After calculating  the  two-spin eigenvalues $\pi_{k}$ ($k=1, 2, 3, 4$) and the   two-spin eigenstates $|\Pi_{k}\rangle$ of   Eq. (\ref{E4} having the representation matrix: $M(t)=\sum_{m}\pi _{m}|\Pi_{m}\rangle \langle \Pi_{m}|$ with $\pi_{m}\geq 0$ and $\sum_{m}\pi _{m}=1$, the LQFI is calculated by using the closed expression \cite{Kim2018,QFI6,QFI7} giving by
\begin{equation*}
F(t)=1-\pi _{R}^{\max },
\end{equation*}%
$\pi _{R}^{\max }$ represents the highest eigenvalue of the symmetric matrix $R=[r_{ij}]$. Based on the Pauli spin-$\frac{1}{2}$ matrices $\sigma ^{i}\,(i=1,2,3)$ and the  elements $\xi_{mn}^{i}=\langle \Pi_{m}|I\otimes \sigma ^{i}|\Pi_{n}\rangle $, the  
symmetric matrix elements $r_{ij}$ are given by
\begin{equation*}
r_{ij}=\sum_{\pi _{m}+\pi _{n}\neq 0}\frac{2\pi _{m}\pi _{n}}{\pi _{m}+\pi
_{n}}\xi_{mn}^{i}(\xi_{nm}^{j})^{\dagger }.
\end{equation*}%
For a two-spin-qubits maximally correlated   state, the   LQFI function has $F(t)=1$. The case of  $0<F(t)<1$ means that  the   states have   partial LQFI's nonlocal correlation. 
\item \textbf{LQU }\newline
 LQU  of Wigner–Yanase (WY) skew information \cite{U0} is   realized to use  as an another type of two spin-qubits' nonlocal correlations   \cite{U0,U1,U2}. For the two spin-qubits' 
density matrix $M(t)$ of Eq. (\ref{E4}), the LQU can be calculated by
  \cite{U1}
\begin{eqnarray}  \label{Eq11}
U(t)&=& 1-\lambda_{max}(\Lambda_{AB}),
\end{eqnarray}
$\lambda_{max}$ designs the largest eigenvalue of the $3\text{x}3$-matrix $\Lambda=[a_{ij}]$, which  have the   elements:
\begin{eqnarray*}  \label{Eq12}
a_{ij}=\text{T}r\mathbf{\big\{}\sqrt{M(t)}(\sigma_{i}\otimes I)\sqrt{M(t)}(\sigma_{j}\otimes I)\mathbf{\big\}}.
\end{eqnarray*}
\item \textbf{Logarithmic negativity (LN)}\\
 We employ the logarithmic negativity \cite{Mn17}   to measure of the generated two-spin-qubits entanglement.
The LN expression is based on the negativity's definition $\mu_{t}$ \cite{Mn17} (which is defined as the absolute sum of the matrix's negative eigenvalues $(M(t))^T$ of the partial transposition of the two spin qubits density matrix $M(t)$ of Eq. (\ref{E4}).   The LN can be expressed as:
\begin{eqnarray}\label{H17}
N(t)&=&\log_{2}[1+2\mu_{t}],
\end{eqnarray}
The $N(t)=0$     for a  disentangled two-spin state,  $N(t)=1$ for   a maximally  entangled two-spin state, and $0\leq N(t) \leq 1$ for   a  partially    entangled two-spin state. 
 \end{itemize}
\section{Two spin-Heisenberg-XYZ-qubits dynamics}\label{SEC-4}
\begin{figure}[h]
\centering
\includegraphics[height=5cm,width=9cm]{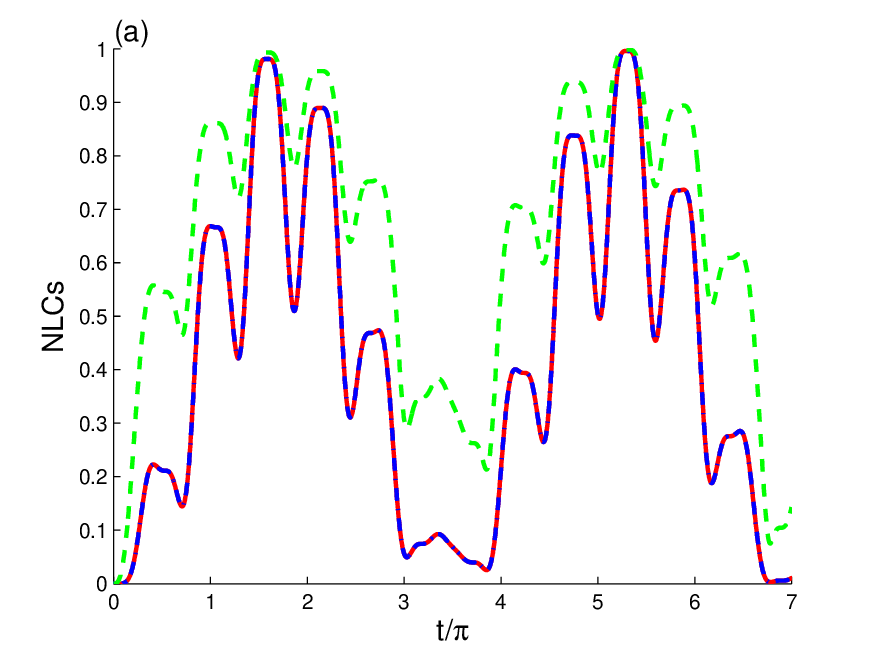}
\includegraphics[height=5cm,width=9cm]{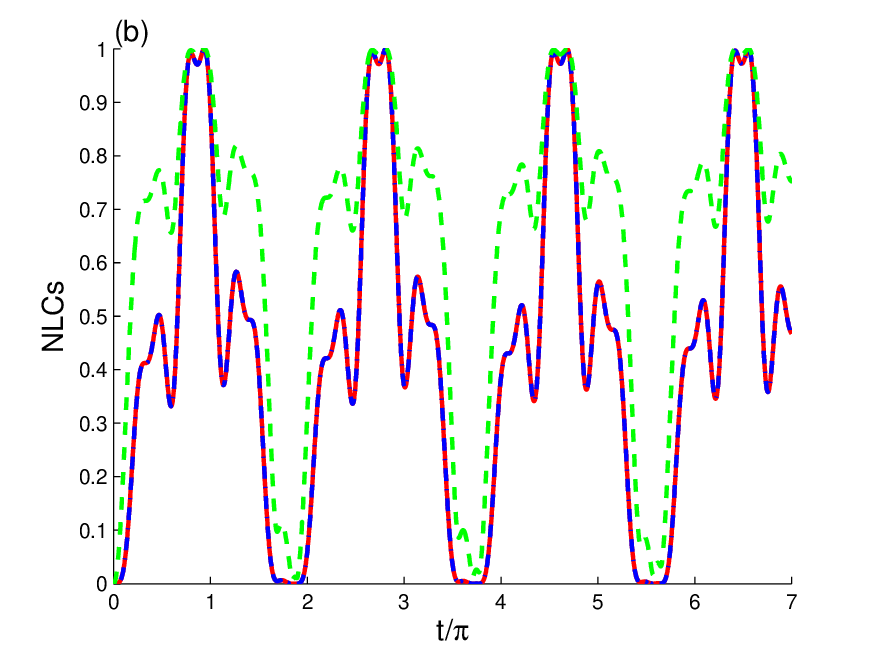} 
\includegraphics[height=5cm,width=9cm]{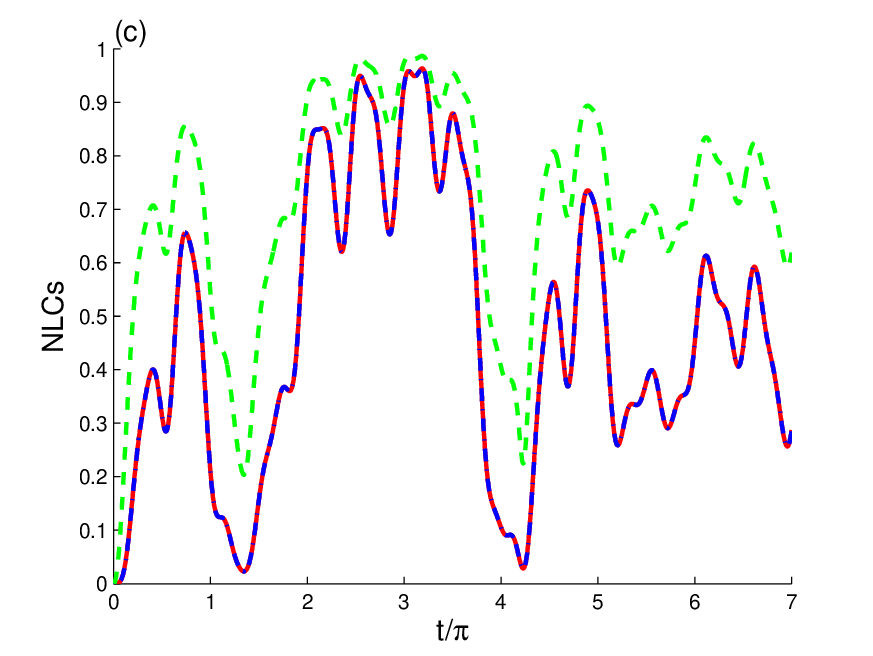}
\caption{The dynamics of the generated local-QFI, local-QU, and log-negativity correlations due to  the   couplings  $(J_{\alpha}, J_{y},  J_{z})=(0.8, 0.8, 0.8)$ are shown under the effects  of the  applied magnetic field  $(B_{m},b_{m})=(0.3, 0.5)$ and the $D_{x,y}$  interactions: $(D_{x}, D_{y}) =(0.0, 0.0)$ in  (a),  $(D_{x}, D_{y}) =(0.5, 0.0)$ in (b), and  $(D_{x}, D_{y}) =( 0.5, 0.5)$ in (c). }
\label{F1}
\end{figure} 
\begin{figure}[h]
\centering
\includegraphics[height=5cm,width=9cm]{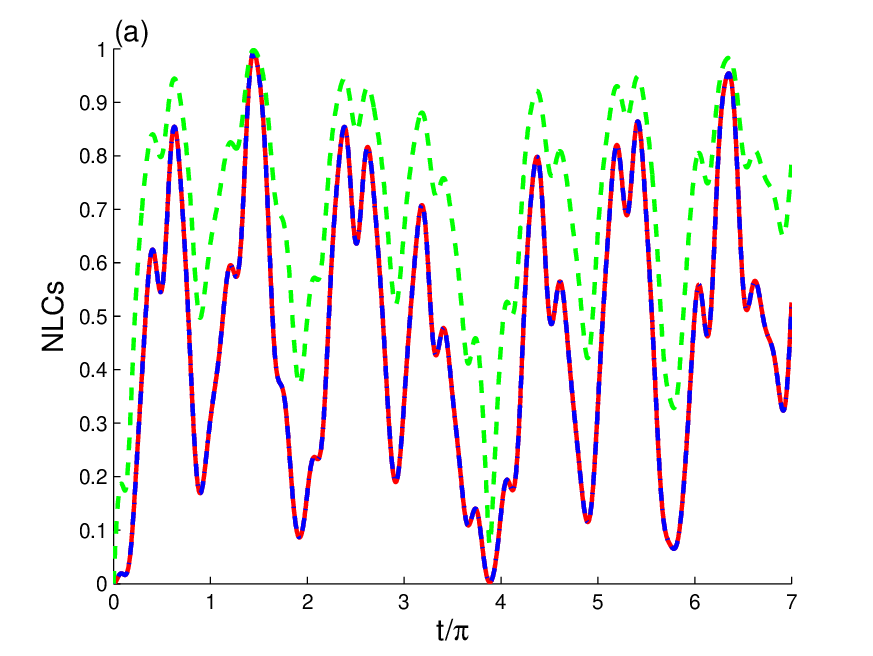} %
\includegraphics[height=5cm,width=9cm]{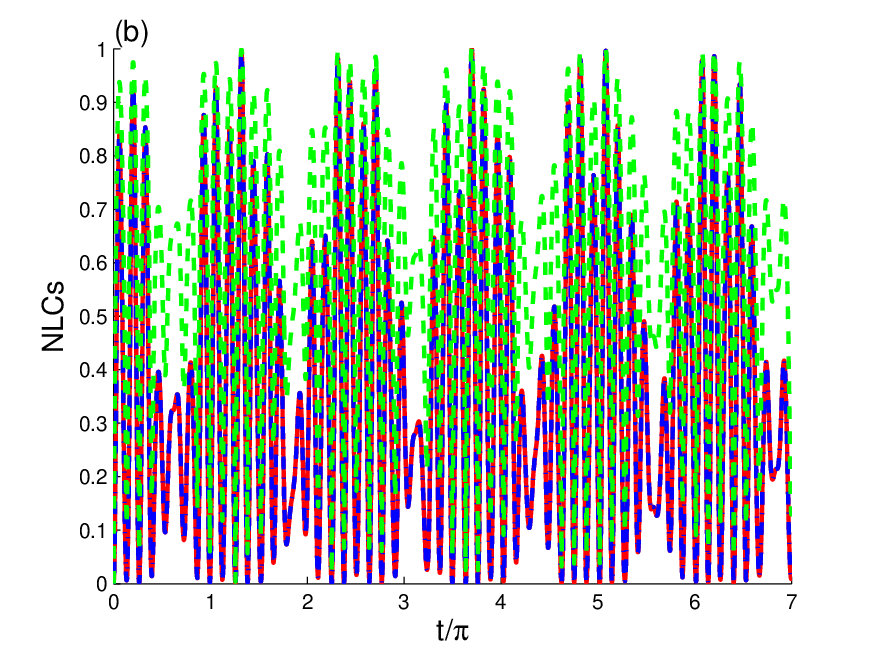}
\includegraphics[height=5cm,width=9cm]{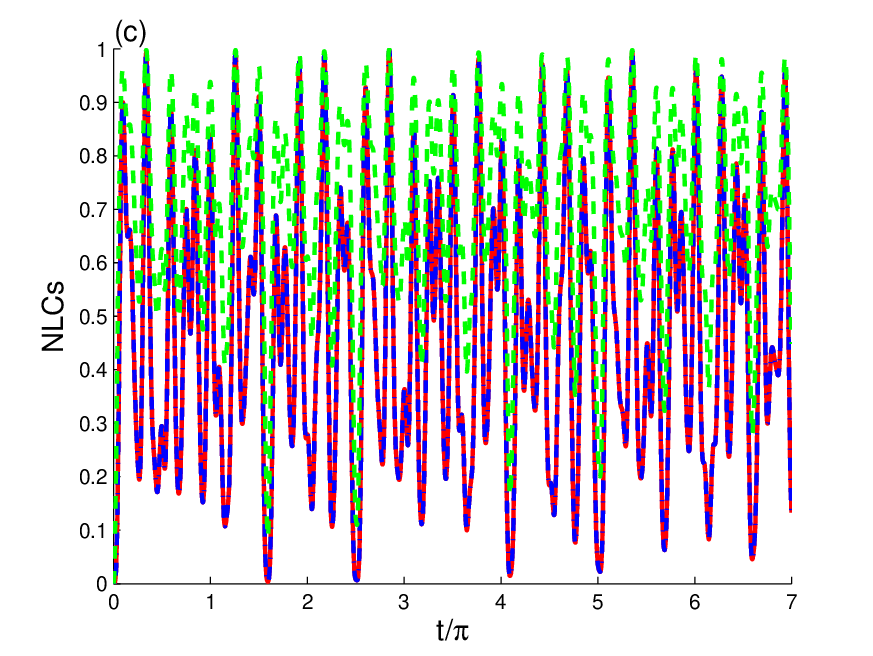}
 \caption{The LQFI (red solid curve), LQU (boule dash-dotted curve), and log-negativity (green dashed curve)  dynamics 
 of Fig.\ref{F1}c  are plotted   for different  Heisenberg-XYZ  couplings: $(J_{x}, J_{y},  J_{z})=(1, 0.5, 1.5)$ in (a) and  $(J_{x}, J_{y},  J_{z})=(5, 1, 1.5)$ in (b), and  strong $x,y$-spin-orbit interactions  $D_{x}=D_{y}=2$ in (c).}
\label{F2}
\end{figure}
\begin{figure}[h]
\centering
\includegraphics[height=5.1cm,width=9cm]{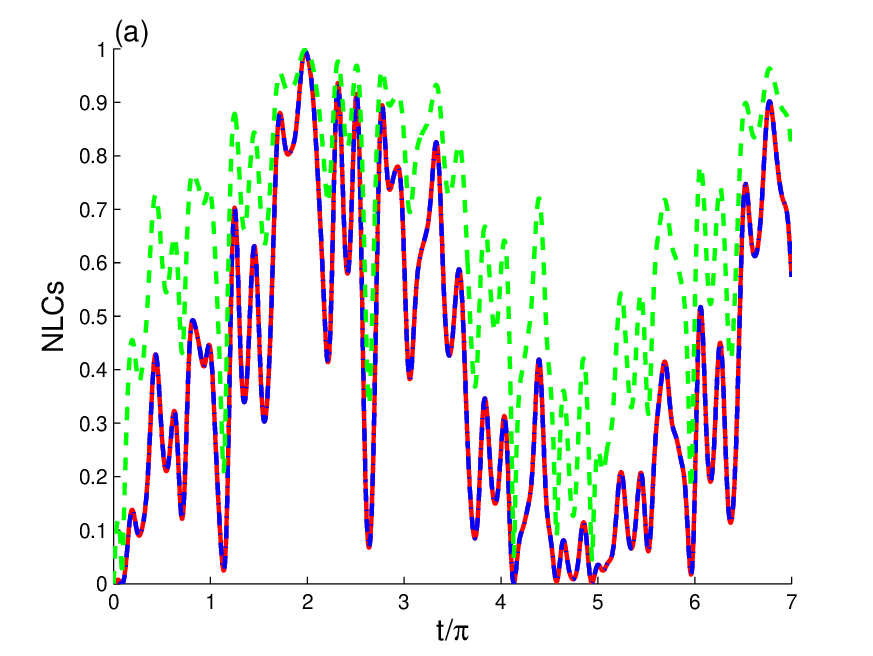} %
\includegraphics[height=5.1cm,width=9cm]{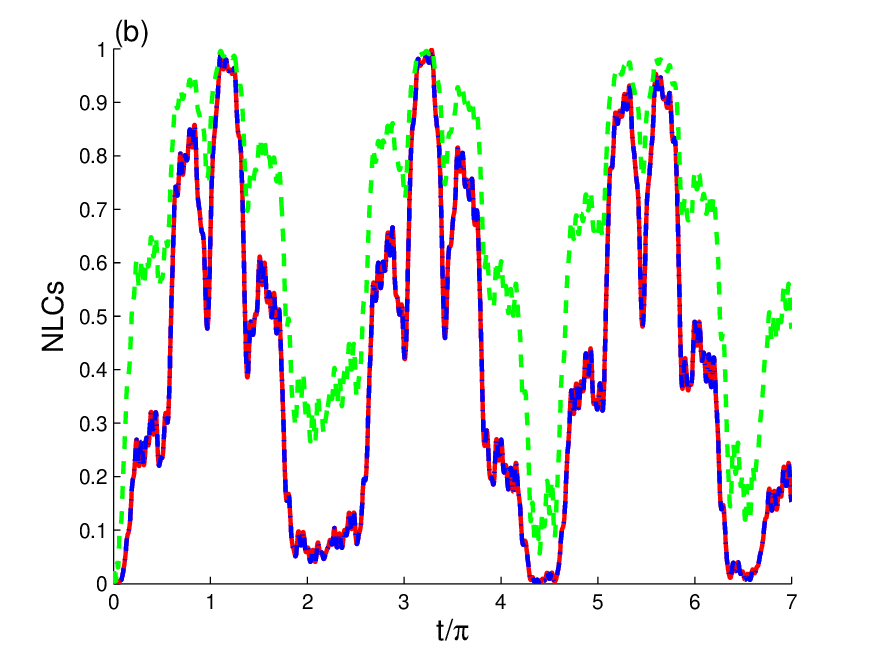}
\caption{The LQFI (red solid curve), LQU (boule dash-dotted curve), and log-negativity (green dashed curve)  dynamics  
 of Fig.\ref{F2}a (for $(J_{x}, J_{y},  J_{z})=(1, 0.5, 1.5)$, $(B_{m},b_{m})=(0.3, 0.5)$, and $D_{x}=D_{y}=0.5$)  are plotted  for different large magnetic-field uniformities:  $B_{m}=2$ in (a) and  $B_{m}=10$ in (b).}
\label{F3}
\end{figure}
\begin{figure}[h]
\centering
\includegraphics[height=5.1cm,width=9cm]{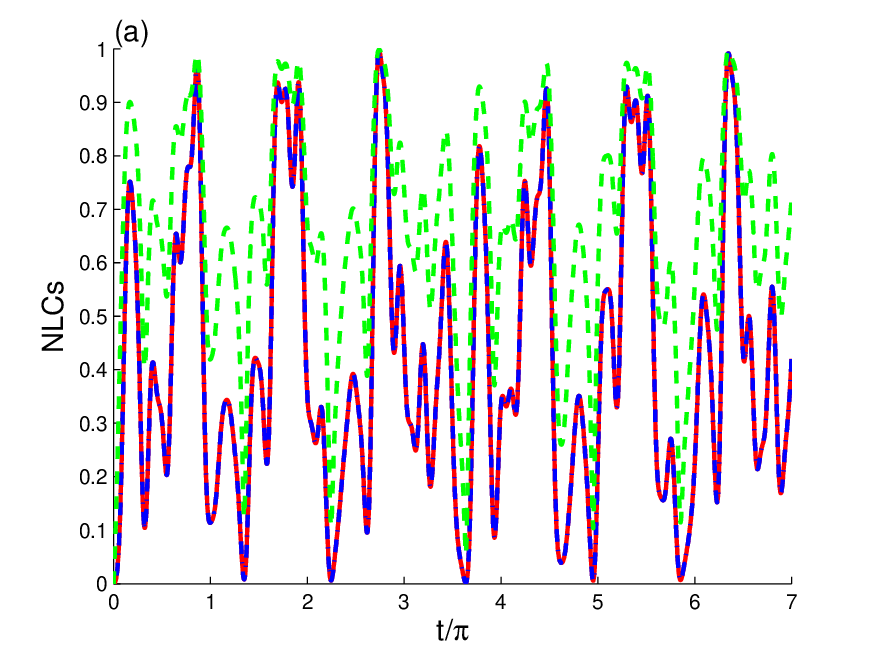} %
\includegraphics[height=5.1cm,width=9cm]{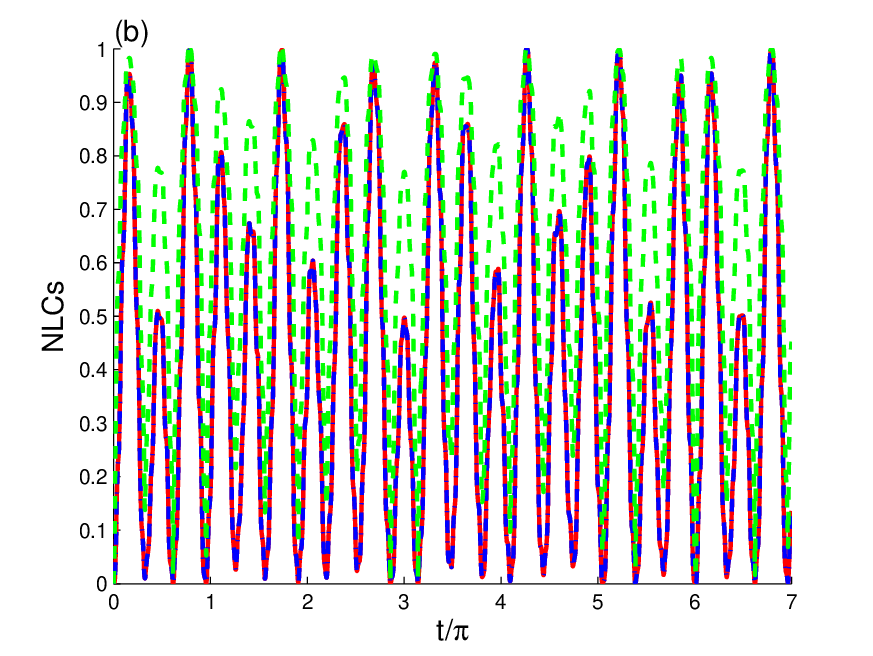}
\caption{The LQFI (red solid curve), LQU (boule dash-dotted curve), and log-negativity (green dashed curve)   dynamics  
 of Fig.\ref{F2}a (for $(J_{x}, J_{y},  J_{z})=(1, 0.5, 1.5)$, $(B_{m},b_{m})=(0.3, 0.5)$, and $D_{x}=D_{y}=0.5$)  are plotted  for different large magnetic-field inhomogeneities:  $b_{m}=2$ in (a) and  $b_{m}=10$ in (b). }
\label{F4}
\end{figure}
 \begin{figure}[h]
\centering
\includegraphics[height=5cm,width=9cm]{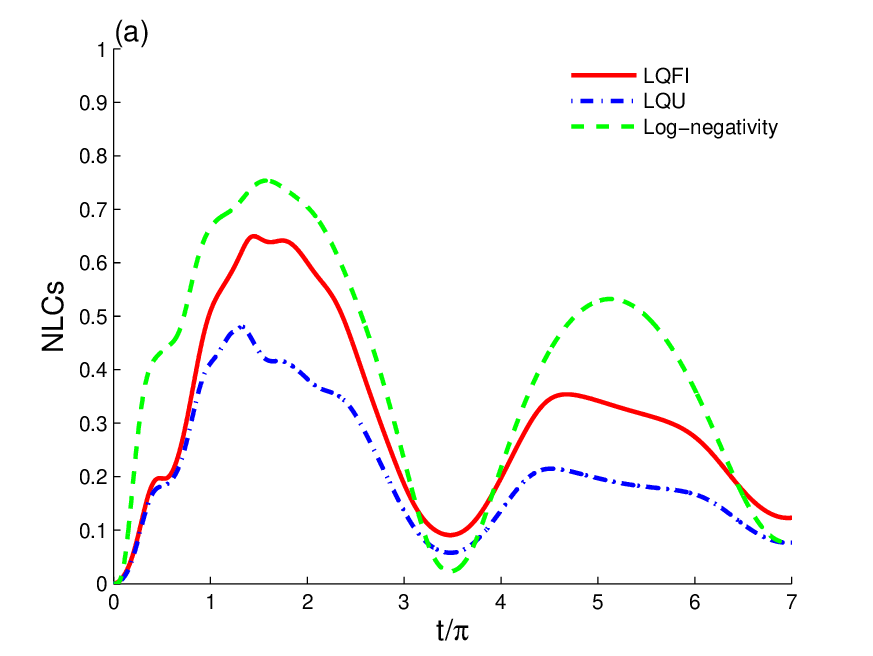}
\includegraphics[height=5cm,width=9cm]{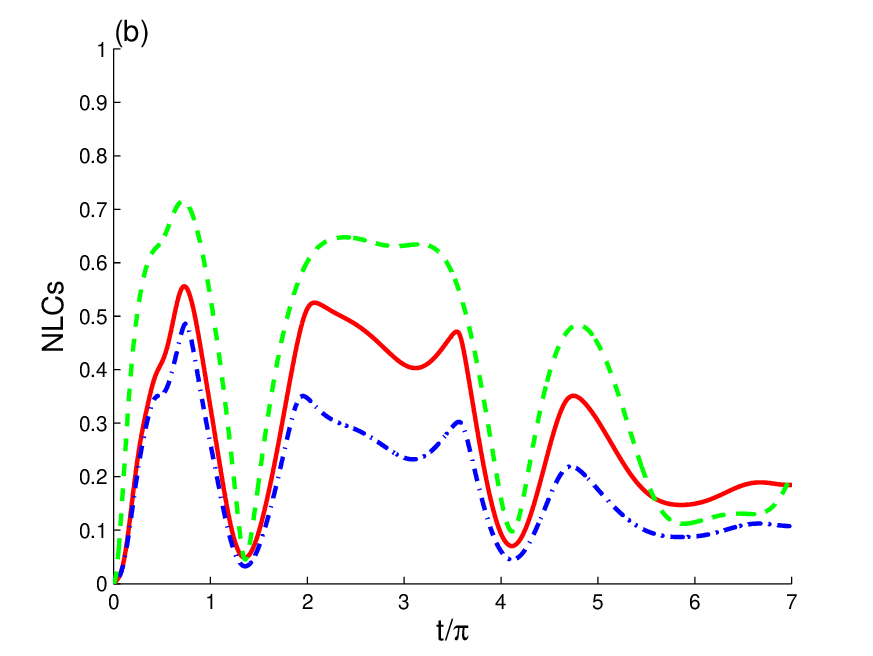} 
\includegraphics[height=5cm,width=9cm]{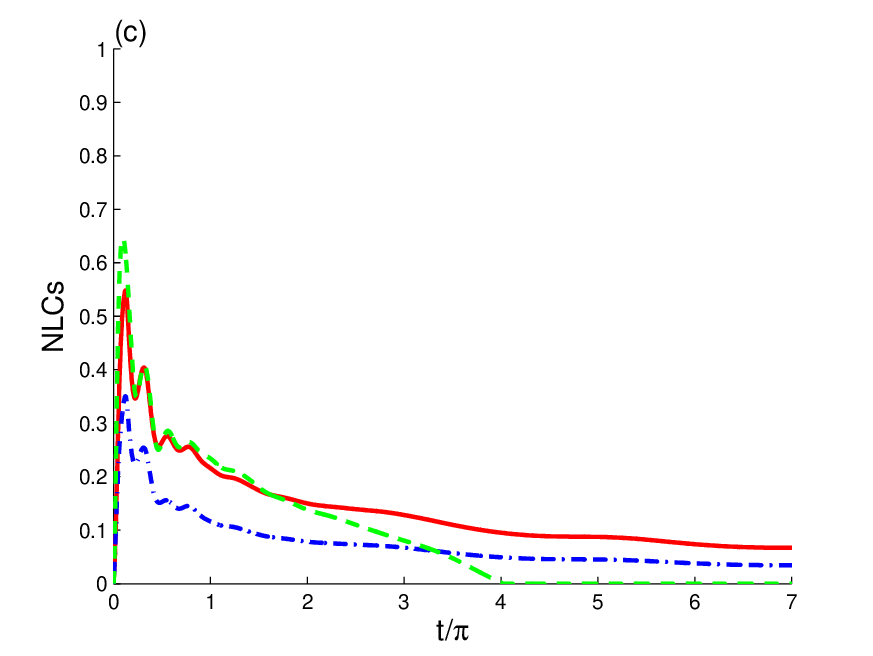}
\caption{The   LQFI (red solid curve), LQU (boule dash-dotted curve), and log-negativity (green dashed curve) dynamics  is shown in the presence of the   ISSD   $\gamma=0.05$  and the   magnetic field  $(B_{m},b_{m})=(0.3, 0.5)$ with the couplings  $J_{\alpha}= 0.8$ for different   couplings: $D_{k}=0 (k=x,y)$ in  (a),  $D_{k}=0.5$ in (b), and  $D_{k}=2$ in (c). }
\label{F5}
\end{figure} 
 \begin{figure}[h]
\centering
\includegraphics[height=5cm,width=9cm]{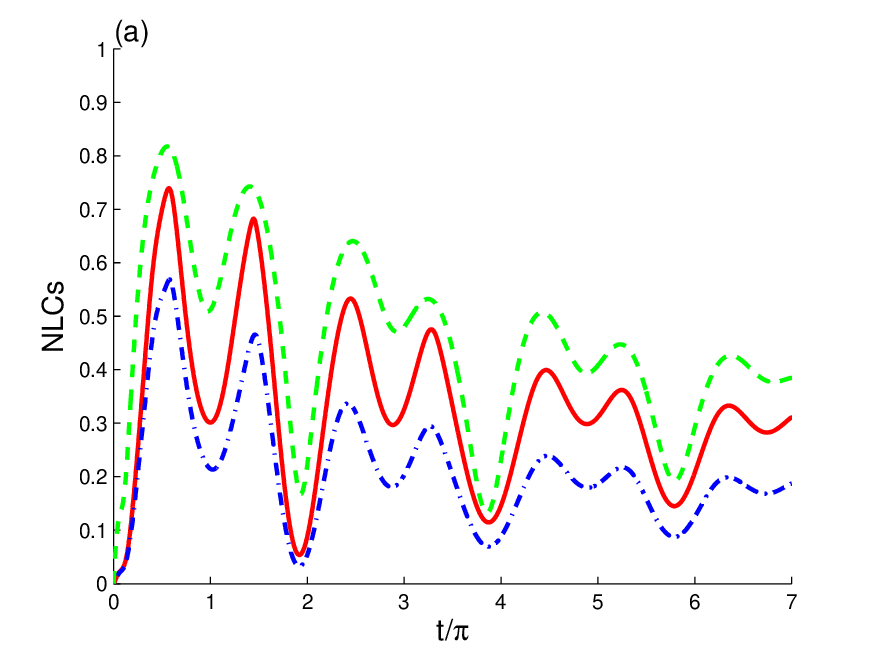}
\includegraphics[height=5cm,width=9cm]{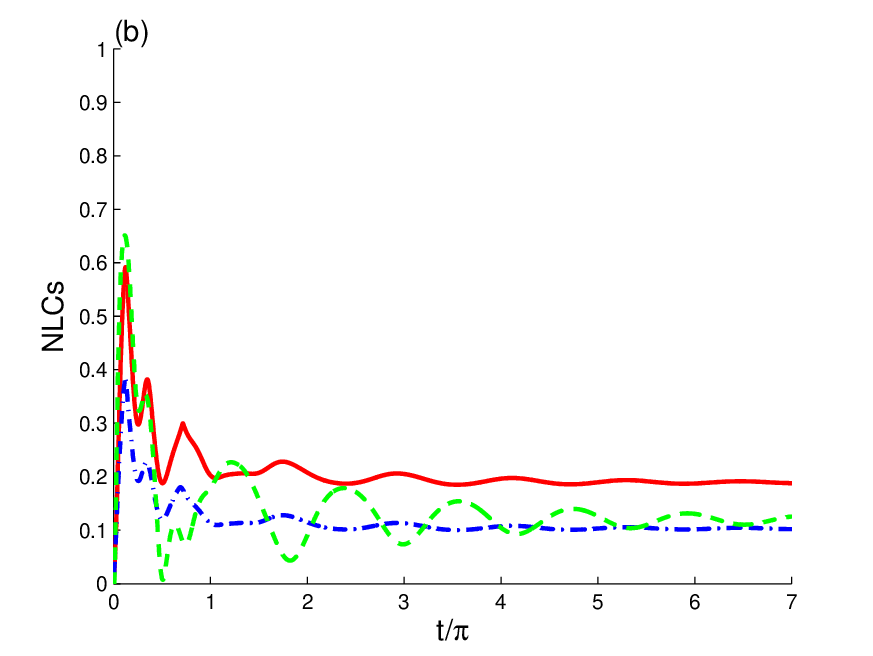} 
\caption{The two spin-qubits correlation dynamics of the  of Fig.\ref{F5}b and c   is shown  but for strong  spin-spin couplings  $(J_{x}, J_{y},  J_{z})=(1, 0.5, 1.5)$. }
\label{F6}
\end{figure} 
 \begin{figure}[h]
\centering
\includegraphics[height=5cm,width=9cm]{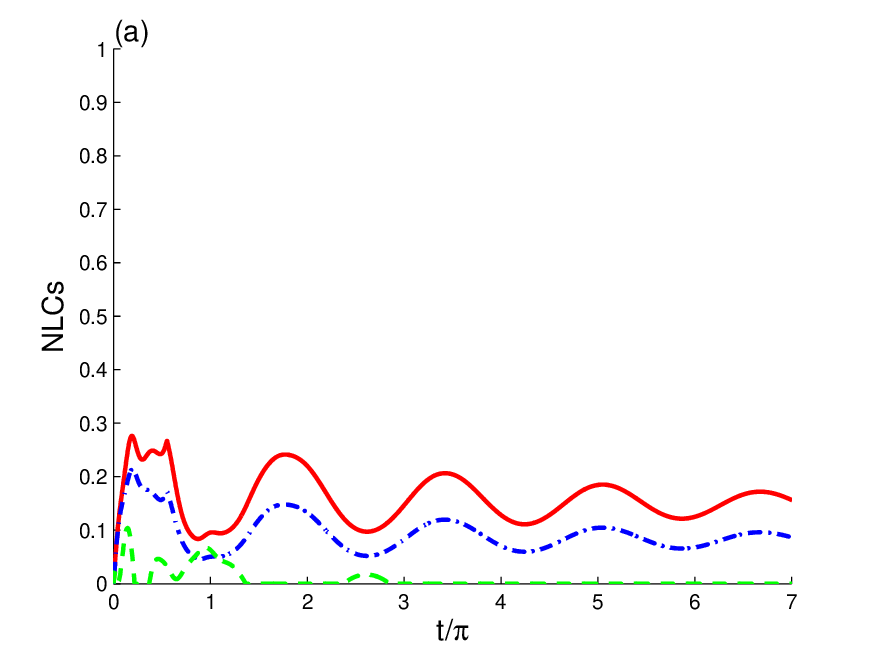}
\includegraphics[height=5cm,width=9cm]{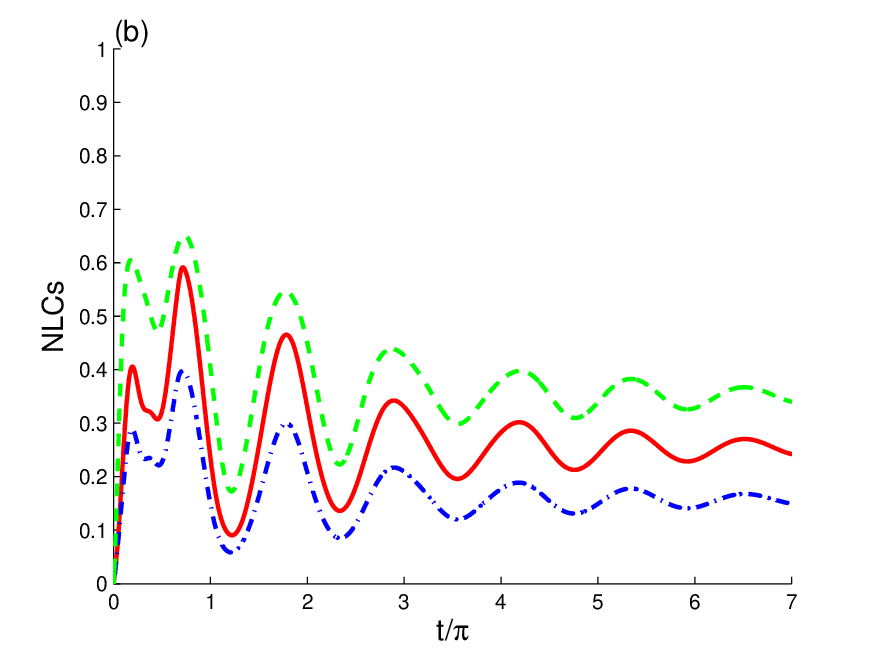} 
\caption{The dynamics of the LQFI (red solid curve), LQU (boule dash-dotted curve), and log-negativity (green dashed curve) of Fig.\ref{F3}a   is shown  but for large   EIMF's uniformity $(B_{m},b_{m})=(2, 0.5)$ in (a) and  large   EIMF's inhomogeneity   $(B_{m},b_{m})=(0.3, 2)$ in (b). }
\label{F7}
\end{figure} 
 Here, we will explore the role of $J_{\alpha}$-spin-spin interactions supported by    $x,y$-spin-orbit interactions in the generation dynamics of the two-qubit non-local correlations (of   LQFI,  local LQU, and LN's entanglement general Heisenberg-XYZ (non-X)-states in the presence of an   $x-$direction EIMF.  
 To explore the generation of the two-spin-qubits non-local correlations, we consider that the two spins are initially in their uncorrelated upper states  $|1_{A}\rangle\otimes|1_{B}\rangle$,  which its density matrix has no nonlocal correlations of the considered quantifiers.
Our focus is on the  $J_{\alpha}$-spin-spin interaction effects  ($D_{x}$ and $D_{y}$), and inhomogeneous $x-$direction magnetic field parameters $(B_{m}$ and $b_{m})$ in the presence of the intrinsic spin decoherence  (ISSD) coupling.
\\ \indent
As our first analysis, we display the dynamics of the two spin qubits nonlocal correlations of the  LQFI, LQU,   and LN, generating due to the couplings  $(J_{x}, J_{y},  J_{z})=(0.8, 0.8, 0.8)$ supported by different intensities of     $x,y$-spin-orbit interactions in the presence of the inhomogeneous $x-$direction magnetic field having weak uniformity and inhomogeneity     $(B_{m},b_{m})=(0.3, 0.5)$.
 In the absence of the intrinsic spin-spin decoherence    $\gamma=0$ and the $x,y$-spin-orbit interactions $(D_{x}, D_{y}) =(0.0, 0.0)$,   the  Fig.\ref{F1}(a) illustrates that the two-spin-qubits LQFI, LQU,   and log-negativity grow to reach their maximum. The nonlocal correlations of the  LQFI, LQU,   and log-negativity undergo slow quasi-regular oscillations having the same frequencies and different amplitudes. LQFI and LQU have the same behavior, i.e., the two-spin-qubits correlation is called the "Fisher-Wigner–Yanase nonlocal correlation". The log-negativity amplitude is always greater than those of the  LQFI and LQU.    
Under these circumstances of  weak coupling regime  of $J_{\alpha}= 0.8$   and the applied inhomogeneous $x-$direction magnetic field (weak  uniformity   and   inhomogeneity), the 
 initial pure-uncorrelated two-spin state undergoes different time-dependent partially correlated states, except particular time, it transforms maximally correlated states.
The two-spin states have maximal correlations of   Fisher-Wigner–Yanase nonlocal correlation ($F(t)=U(t)=1$) and log-negativity $N(t)=1$ at the same time. 
At particular times, we observe that partially two-spin entangled states have no LQFI or  LQU correlation. 
\\ \indent
 The effects of weak intensities of     $x,y$-spin-orbit interactions are shown in Fig.\ref{F1}(b). As is clear in this figure, the regularity and fluctuations of the generated  Fisher-Wigner–Yanase and log-negativity nonlocal correlations are substantially more than previously presented in the absence of the $x,y$-spin-orbit interaction.
The weak $D_{x}$-spin-orbit interaction  $(D_{x}, D_{y}) =( 0.5, 0)$ dramatically improves the appearance of the intervals of the maximal Fisher-Wigner–Yanase and log-negativity nonlocal correlations, as well as the intervals in which two-spin entangled states have no LQFI or  LQU correlation. 
 In Fig.\ref{F1}(c), we combined the $D_{x}-$ and $D_{y}-$spin-orbit interactions  $(D_{x}, D_{y}) =( 0.5, 0.5)$ into $x,y$-spin-orbit interactions. As is clear in this figure, the  NLC  fluctuations between their partial and maximal values are
substantially fewer than previous results in Figs.\ref{F1}(a,b). Furthermore, the NLC frequency has been reduced while the lower bounds of the  Fisher-Wigner–Yanase and log-negativity nonlocal correlations are shifted up. This means that the combined $D_{x}-$ and $D_{y}-$spin-orbit interactions  $(D_{x}, D_{y}) =( 0.5, 0.5)$  improves the generated partial two-spin-qubits Fisher-Wigner–Yanase and log-negativity   correlations. 
\\ \indent
 Fig.\ref{F2} shows that the higher couplings of $J_{\alpha}$-spin-spin interactions and  $x,y$-spin-orbit interactions ($(J_{x}, J_{y},  J_{z})=(1, 0.5, 1.5)$ in (a) and  $(J_{x}, J_{y},  J_{z})=(5, 1, 1.5)$ in (b), and  strong $D_{x,y}$-spin-orbit interaction  $D_{x}=D_{y}=2$ in (c)) have a high ability to  enhancing the  arisen  two-spin-qubits' NLC of the Fisher-Wigner–Yanase and log-negativity.
 By comparing the generated spin-spin NLCs showing in  Figs.\ref{F1}(c) and   \ref{F2}(a), we find that the relative strong couplings  of $J_{\alpha}$-spin-spin interactions $(J_{x}, J_{y},  J_{z})=(1, 0.5, 1.5)$ (which are supported by a weak $D_{x,y}$-spin-orbit interactions $(D_{x}, D_{y}) =( 0.5, 0.5)$) increases the amplitudes and  frequencies of the      
Fisher-Wigner–Yanase and log-negativity's oscillations. Fig.\ref{F2} shows that the higher $J_{\alpha}$-couplings lead to the spin-spin NLCs' oscillations have  more regularity and fluctuations. The time positions of the maximal Fisher-Wigner–Yanase and log-negativity   correlations are enhanced. 
Fig.\ref{F2}(c) is plotted to see the capability the increase of the $x,y$-spin-orbit interactions $D_{x}=D_{y}=2$ supporting by weak spin-spin interactions $J_{\alpha}= 0.8)$  to enhance the generated spin-spin NLCs when the external magnetic field applied with weak determinants     $(B_{m},b_{m})=(0.3, 0.5)$.
By comparing the qualitative dynamics of the generated Fisher-Wigner–Yanase and log-negativity correlations shown in Fig.\ref{F1}c ($D_{x}=D_{y}=0.5$)  with that shown in  Fig.\ref{F2}c ($D_{x}=D_{y}=2$), we can deduce that the $D_{x,y}$-spin-orbit interactions have a high role in enhancement the generated Fisher-Wigner–Yanase and log-negativity correlations, their amplitudes are increased and their oscillations have more fluctuations between their extreme values.
 In addition, the strong  $x,y$-spin-orbit interactions potentially strength and speed the generation of the Fisher-Wigner–Yanase and log-negativity correlations, due to the $J_{\alpha}$-spin-spin interactions.
\\ \indent
 Figures \ref{F3}  show  Fisher-Wigner–Yanase and log-negativity nonlocal correlation dynamics of Fig.\ref{F2}a (where $(J_{x}, J_{y},  J_{z})=(1, 0.5, 1.5)$, $b_{m}=0.5$, and $D_{x}=D_{y}=0.5$)  are plotted for different uniformities of the applied  EIMF.  
Fig. \ref{F3}a  shows the generated Fisher-Wigner–Yanase and log-negativity   correlations, due to   the $J_{\alpha}$-spin-spin  and  $x,y$-spin-orbit interactions, previously presented  after      applying  an external    magnetic field  (having  a small 
inhomogeneity  $b_{m}=0.5$ and a large uniformity  $B_{f}=2$).
In this case, the increase of the EIMF uniformity delays the growth of the LQFI, LQU as well as log-negativity. It increases the two-spin state's fluctuations between different partially and maximally correlated states.   
The  generations of the Fisher-Wigner–Yanase and log-negativity   correlations shown in Fig.\ref{F2}a (with   $B_{m}=0.5$) and   Fig.\ref{F3}a (with   $B_{m}=2$)  with those shown in  Fig.\ref{F3}b (with   $B_{m}=10$) confirm   that the
increase of the EIMF uniformity will enhance the ability of the strong $J_{\alpha}$-spin-spin interactions supported by weak $x,y$-spin-orbit interactions to create partially and maximally correlated states with more stability; however,  the generated spin-spin NLCs are more sensitive to the EIMF uniformity. 
\\ \indent
In the forthcoming analysis of Fig.\ref{F4},   we keep the system with the   same parameters' values of Fig.\ref{F2}a (where $(J_{x}, J_{y},  J_{z})=(1, 0.5, 1.5)$, $B_{m}=0.3$, and $D_{x}=D_{y}=0.5$)  and consider    different magnetic-field inhomogeneities:  $b_{m}=2$ in (a) and  $b_{m}=10$ in (b). 
 In this case of  Fig.\ref{F4}a,  we notice that the larger EIMF uniformities  enhance  the efficiency of the generation of the Fisher-Wigner–Yanase and log-negativity   correlations.  The EIMF uniformity  increases the two-spin state's fluctuations between different partially and maximally correlated states.  
 The time positions of the maxima  ($F(t)=U(t)=N(t)\approx1$ and minima (zero-value)  ($F(t)=U(t)=N(t)\approx0$ of the generated Fisher-Wigner–Yanase and log-negativity   correlations are enhanced.
 In the case of $(J_{x}, J_{y},  J_{z})=(1, 0.5, 1.5)$, the increase of the EIMF inhomogeneity  have a high role in enhancement the  generated two-spin-qubits' NLCs. Where   NLCs oscillations' amplitudes and fluctuations are increased (see Fig.\ref{F4}b).
\\ \indent
 The next illustration of Figs.\ref{F5}-\ref{F7} is obtained   to show the nonlocal correlation dynamics of the    LQFI, LQU,   and log-negativity  in the presence of the non-zero ISSD  coupling $\gamma=0.05$. By comparing the results of Fig.\ref{F1}a ($\gamma=0.0$) with these of Fig.\ref{F5}a ($\gamma=0.05$),  we find that the LQFI, LQU,   and log-negativity  shows a decaying   oscillatory  dynamical evolutions.
The generations  of the  Heisenberg-XYZ (non-X)-states'NLCs (due to $J_{\alpha}= 0.8$ spin-spin couplings the  applied magnetic field  $(B_{m},b_{m})=(0.3, 0.5)$ without spin-orbit interaction) are weaken and have  different amplitudes (which  are decreased by increasing the ISSD  coupling). After particular interval time, with non-zero ISSD  coupling, the  LQFI and  LQU present different  nonlocal correlations  having different amplitudes with  the same behaviors. 
 Moreover, the  NLCs' robustness (against   the ISSD effect) of the LQFI    and log-negativity  is more than that of the LQU.    
\\ \indent
 As shown in Figs.\ref{F5}b and c, the  increase of the intensities of     $x,y$-spin-orbit interactions ($D_{k}=0 (k=x,y)$ in  (a),  $D_{k}=0.5$ in (b), and  $D_{k}=2$ in (c)) reduce the    NLCs' robustness (against   the ISSD effect) of the LQFI,  LQU    and log-negativity  correlation, the NLCs' amplitudes significantly decrease as the $x,y$-spin-orbit interactions increase.  Moreover, LQFI and  LQU   display
  sudden changes  at different times. The   sudden-changes phenomenon   has been studied  theoretically   \cite{SCH1} and   experimentally  \cite{SCH4} (see Figs.\ref{F5}b and c).  
 For very strong   $x,y$-spin-orbit interactions  $D_{k}=2  $ (see Fig.\ref{F5}c), we observe that the two-spin-qubits log-negativity  drops instantly to zero   at  a particular time  for a long time (sudden-death  LN-entanglement phenomenon), then the disentangled two-spin states have only different stable NLCs of the  LQFI and   LQU.
 We can deduce that the NLCs' decay resulting from ISSD   can be enhanced  by increasing the  intensities of     $x,y$-spin-orbit interactions.
\\ \indent
 By comparing the Figs.\ref{F5}(b,c) and Figs.\ref{F6}(b,c), we find that, the  strong  spin-spin couplings  $(J_{x}, J_{y},  J_{z})=(1, 0.5, 1.5)$ reduce the  ISSD effect and improve the NLCs' robustness (against   the ISSD effect) of the LQFI,  LQU, and  LN. For very strong   $x,y$-spin-orbit interactions $D_{k}=2$ (see Fig.\ref{F6}b), the sudden-death  LN-entanglement phenomenon does not occur, except  at the time $t\approx0.5\pi$ it occurs instantaneously. The generated two-spin states have  different stable partial NLCs of the  LQFI,  LQU, and  LN. 
  In this case, the NLCs' decay resulting from ISSD   can be weakened   by strengthening   the  spin-spin interactions.
  \\ \indent
In the presence of the   ISSD  effect $\gamma=0.05$, Fig.\ref{F7} shows   the generated NLCs  of the Fig.\ref{F3}a  (or  it shows the degradation of NLCs  of  Fig.\ref{F6}a)  after  strengthening  the   EIMF's uniformity $(B_{m},b_{m})=(2, 0.5)$ in (a) and    EIMF's inhomogeneity    $(B_{m},b_{m})=(0.3, 2)$ in (b). From Fig.\ref{F7}a, we observe that the large EIMF's uniformity $B_{m}=2$ increases the NLCs' decay resulting from ISSD. The time intervals in which the disentangled two-spin states have only different stable NLCs of the  LQFI and   LQU appeared.   Moreover, the  NLCs' robustness (against the ISSD effect) of the LQFI    and LN is reduced by increasing the large EIMF's uniformity.
The outcomes of Fig.\ref{F7}b   illustrate that strengthening the  EIMF's inhomogeneity    $b_{m}=2$ also increases the degradation of the  NLC  functions.
In this case of the parameters:  $b_{m}=2$, $(J_{x}, J_{y},  J_{z})=(1, 0.5, 1.5)$,  and  $D_{k}=0.5$, we observe that: (1)  the generated NLCs  (LQFI,  LQU    and entanglement) of the Fig.\ref{F3}a  degrade (due to and  ISSD  effect) and reach their partial stable oscillatory behaviors, quickly,  comparing with the case where the small value of  $ b_{m}= 0.5$ of Fig.\ref{F6}a. 
We find that the ability of the EIMF's inhomogeneity to enhance the     ISSD  effect is small compared to that of the  EIMF's uniformity.
\section{Conclusion}

\label{SEC-5}
In this investigation, the Milburn intrinsic decoherence model and Heisenberg XYZ model are used to examine the embedded capabilities in spin-spin interaction and spin-orbit interaction (that describes   $x,y$-DM interactions)   to generate nonlocal correlations (realizing by LQFI,  LQU, and  LN) of general two-spin-qubits (non-X)-states under the effects of the  EIMF's uniformity and the inhomogeneity.
In the presence and absence of the ISSD,  the dependence of the generated nonlocal correlations on the parameters, of spin-spin interaction and spin-orbit interactions as well as of the EIMF's uniformity and the inhomogeneity, are explored.    
It is found that the spin-spin  Heisenberg XYZ and  $x,y$-spin-orbit interactions have a high capability to raise non-local correlations in the presence of a weak external magnetic field.  The spin-orbit interactions have a high role in the enhancement of the generated two-spin-qubits Fisher-Wigner–Yanase and log-negativity correlations,    their oscillations' amplitudes and fluctuations are increased.   
 In the presence and absence of the ISSD, the NLCs' generations are weakened and have different amplitudes,  decreasing by increasing the ISSD  coupling. The  NLCs' robustness (against the ISSD effect) of the LQFI    and log-negativity is more than that of the LQU.
The phenomenon of the sudden changes occurs during the LQU and LQFI dynamics whereas the sudden death occurs during log-negativity-entanglement dynamics. 
 The NLCs' decay resulting from ISSD   can be enhanced by increasing the intensities of     $x,y$-spin-orbit interactions.
 Strengthening the spin-spin interactions weakens the NLCs' decay resulting from ISSD. The generated NLCs degrade (due to an ISSD  effect with a large IMF's inhomogeneity) and reach their partially stable oscillatory behaviors, quickly. 
The ability of the IMF's inhomogeneity to increase the     ISSD  effect is small compared to that of the  EIMF's uniformity.

\end{document}